\documentclass[aps,a4paper,%12pt,rpsfig
,nofootinbib]{revtex4}
\usepackage{amssymb,latexsym}
\usepackage{graphicx}
\usepackage{bm}
\usepackage{mathrsfs}
\usepackage{xcolor,color,graphicx,graphics}
\usepackage[all]{xy}
\usepackage{epsfig,subfigure}
\usepackage{latexsym,amssymb,amsmath,amsfonts}
\usepackage[english]{babel}
\usepackage[OT1]{fontenc}
\usepackage[latin1]{inputenc}
\usepackage{makeidx}
\usepackage{hyperref}
\usepackage{color,graphicx,graphics,wrapfig,epsf}
\begin{document}

\title{About some compression algorithms}
\author{Orchidea Maria Lecian\\
\small
lecian@icra.it\\
ICRA c/o Physics Department, Sapienza University of Rome,\\ Piazzale Aldo Moro, 5- 00185 Rome, Italy;\\
Faculty of Civil and Industrial Engineering, Sapienza University of Rome,\\ Via Eudossiana, 18- 00184 Rome, Italy;\\
Faculty of Information Engineering, Sapienza University of Rome,\\ Via Eudossiana, 18- 00184 Rome, Italy;\\
\normalsize Brunello B. Tirozzi\\
\small brunellotirozzi@gmail.com\\
Department of Physics,\\
\small
University La Sapienza of Rome, Italy.\normalsize}
\date{June 2023}

%\begin{document}

\maketitle

\section*{Abstract} We use neural network algorithms for finding compression methods of images in the framework of iterated function systems which is a collection of the transformations of the interval $(0, 1)$ satisfying suitable properties.
\section*{Key-words:} Fractal compression; neural networks; Hertz method.
\section{Introduction\label{sectionaa}} Be the series $x_1, ..., x_n$. A function $g$ has to be found, such that
\begin{equation}\label{v31}
x_{t+1}=g(x_t).
\end{equation}
The series can be the trajectory of a dynamical system or the realisation of some stochastic process, where the latter can be obtained after experimental evidence.\\ Two approaches are going to be discussed.\\
The auto-regressive integrated moving average (ARIMA) models are already apt to provide with a solution of this problem; nevertheless there is the limitation to their application since they are valid for stationary processes only, and it is sometimes necessary to implement many prime differences to obtain from the series the initial data of a stationary succession, which peculiarity augments the characteristic dimension of of the process artificially. The neural networks can be applied to any succession without the need of the property of stationarity; they are generally used to reconstruct any series of data from any non-linear non-deterministic transformation.\\
The paper is organised as follows.\\
In Section \ref{sectionaa}, the possibility to apply neural networks to reconstruct series of data form non-linear non-deterministic transformations is presented.\\
In Section \ref{section2}, the Hertz method is analysed.\\
In Section \ref{section3}, the partition of the phase-space methodology is put forward.\\
In Section \ref{section4}, patterns-learning neural networks are exposed.\\
In Section \ref{section5}, the Hertz method is further investigated.\\
In Section \ref{section6}, layers networks are exhibited.\\
In Section \ref{section7}, the images-compression methods are set.\\
In Section \ref{section8}, the contractive transformations are further characterised.\\
In Section \ref{section9}, the qualities of the winner neuron are revelaed.\\
In Section \ref{section10}, prospective studies are envisaged.
\section{The Hertz method\label{section2}} As an example, the Hertz reconstruction method \cite{ref13} allows one to determine with much precision the graphic of the chaotic transformation of the interval $(0, 1)$ in itself,i.e.
\begin{equation}
x_{n+1}=4x_n(1-x_n),\ \ x_n\in(0, 1).
\end{equation}
In general, the data to be analysed will be apt to be grouped in $d$-dimensional vectors, for which the block of the first $d$ data can be considered as the components of the first vector, the second block of $d$ data can be considered as the components of the second vector, and so on, according to the following scheme:
\begin{subequations}
\begin{align}
&\underline{x}=(x^1_m), \ \ m=1, ..., d,\\
&x^k_m=x_{kd+m}, \ \ m=1, ..., d,
\end{align}
\end{subequations}
i.e. such that $d$ is the characteristic dimension of the process. In the $1$-dimensional case here considered, $d$ can be identified with the correlation length between the data which is given from the smallest $k$ such that
\begin{equation}\nonumber
Ex_tx_{t+k}=(Ex_t)^2.
\end{equation}
If one has also a geometric structure, data have to be considered, which have dimension equal to that of the MANIFOLD to which they belong; afterwords, the implementation in the above can be applied. In general, on has thus to consider  a transformation
\begin{equation}\label{v33}
\underline{x}^{k+1}=\underline{g}(\underline{x}^k), \ \ \underline{x}^k\in \mathbb{R}^m.
\end{equation}
\section{The Hertz method: the partition of the phase-space methodology\label{section3}}
For the sake of simplicity, one can consider the case $m=1$; it is worth recalling the method of Hertz et al. \cite{ref13}.\\
The method consists in establishing a partition of the state space $X$, which is taken as the same of the interval $(0, 1)$ for the sake of directness, where the partition consists of intervals (blocks) $B_{\alpha, a}=1, ..., M$; in other words,
\begin{equation}\label{v34}
X=\cup_{\alpha=1}^{\alpha=M}B_{\alpha},\ \ B_\alpha\cap B_\beta=0\ \ \forall \alpha\neq\beta.
\end{equation}
The dimension of these blocks is not necessarily constant, as one can learn from the recursive process used for their determination, which is illustrated in the below.\\
For each block $B_\alpha$, the mean value $x_\alpha$ is defined, of the data which belong to the block; the dispersion $\sigma^2_\alpha$ of the is defined as well as
\begin{equation}\label{v35}
x_\alpha=\frac{1}{N_\alpha}\sum_{x\in B_\alpha}x\sigma_\alpha^2=\frac{1}{N_\alpha}\sum_{x\in B_\alpha}(x-x_\alpha)^2,
\end{equation}
where $N_\alpha$ is the number of data which belong to the block $B_\alpha$. The structure $B_1, ..., B_M$ is determined as follows.\\
As a first step, the whole space $X$ is divided into two blocks $B_L$ and $B_R$, where the condition is imposed, that the sum
\begin{equation}
\sigma_{B_L}^2+\sigma_{B_R}^2
\end{equation}
is the least possible one.\\
within this construction, it is fundamental that each block contain a number of points which is greater than a certain quantity.\\
As an example, n the case of data from geology, the condition $N_\alpha>16$ has to be imposed \cite{ref10}: differently, it is impossible to reconstruct the series. This condition, indeed, allows one to avoid the occurrence of extreme situations, such as the possibility that the leftmost block $B_L$ contains only one datum $N_L=1$, and that all the other data belong to the rightmost block $N_R=N-1$. The lower limit of the $N_\alpha$ is determined also after the request that the introduction of the quantities $x_\alpha$ and $\sigma_\alpha^2$ be of interest.\\
This construction is repeated in each of the blocks $B_L$ and $B_R$ until the values of $N_\alpha$ reach the lower limit; finally, a certain number of blocks is constructed, each containing a certain group of data. The presented reconstruction method consists in the approximation of $g$ in each block with an opportune constant, i.e. in finding a local approximation of $g$ in each $B_\alpha$.\\
For these purposes, for each $B_\alpha$ a characteristic function $f_\alpha(x)$ of the block is here newly defined as 
\begin{equation}\label{v37}
f_\alpha(x)=e^{-\frac{(x-x_\alpha)^2}{2\sigma_\alpha^2}};\ \ P_\alpha=\frac{f_\alpha(x)}{\sum_{\alpha=1}^{\alpha=M}f_\alpha(x)}.
\end{equation}
In Eq. (\ref{v37}), the function $P_\alpha$ is obviously defined to acquire the approximate value of $1$ for $x\in B_\alpha$,as the other functions $f_\beta(x)$
are almost vanishing for $x\notin B_\beta$; it is therefore possible to demonstrate straightforward that, in Eq. (\ref{v37}), the most relevant contribution in the denominator is $f_\alpha(x)$ for $x\in B_\alpha$,i.e. from which it is straightforward demonstrated that the fraction equals $1$. $\Box$\\
The procedure is most effective the smaller the $\sigma_\gamma^2$: for this reason, the construction of the blocks $B_\alpha$ is accomplished under the request that the condition of minimum hold. The function $g(x)$ is approximated by a function $f(x)$ which is almost constant in the blocks $B_\alpha$, and which takes the value $f^\alpha$ in such a block as
\begin{equation}\label{v38}
f(x)=\sum_{\alpha=1}^{\alpha^M}f^\alpha P_\alpha(x).
\end{equation}
The constants $f^\alpha$ are determined by imposing the minimum of the Error function
\begin{equation}\label{v39}
E=\sum_{i=1}^{i=N-1}(x_{+1}-f(x_i))^2.
\end{equation}
As before, the minimum can be found by the Monte-Carlo method by identifying $E( \{ f^\alpha \} )$ with the energy of the system.\\
\section{Patterns-learning neural networks\label{section4}} Furthermore, it is interesting to consider the present algorithm also from a different point of view, i.e. as a particular neural network, which learns the patterns $x_1, ..., x_n$.\\
It is the present purpose to newly demonstrate the important concept which s at the basis of all the neural networks applied to the analysis and to the recognition of the data.\\
The following neural structure is considered. \\
The value $x$ assigned to the input neuron is passed to the intermediate units $1, ..., M$. The unit $\alpha$ il qualified after a non-linear output function $P_\alpha(x)$ defined in Eq. (\ref{v37}); $f^1, ..., f^M$ are the weights which connect the output units with the neurons $1, ..., M$. The output neuron receives a quantity
\begin{equation}\label{v310}
f(x)=\sum_{\alpha}f^\alpha P_\alpha(x).
\end{equation}
According to the neural language, the network learns the patterns $(x_i, x_{i+1})$ by varying the weights $f^\alpha$ (where the evolution of the weights corresponds, in this language, to the learning process), such that the difference between the wanted output $x_{i+1}$ in correspondence of the input $x_i$ and the effective output $f(x_i)$ be the least; this procedure corresponds to the calculation of the minimum of Eq. (\ref{v39}). The law of evolution of the weights can be given after the probabilistic change of the Monte-Carlo method, or by the gradient method
\begin{equation}
\Delta f^\alpha=-\eta(k)\frac{\partial E}{\partial f^\alpha},
\end{equation}
where $\eta(k)$ is a coefficient decaying with $k$, $k$ being the number of iterations of the learning process, and $\Delta f^\alpha$ is the variation of the weight $f^\alpha$. The reason of the decreasing behaviour of $\eta(k)$ is explained after considering $E$ as a quadratic function of the variable $x$. If $x$ is on the left of the minimum, it undergoes a positive increase, while if $x$ is on the right of the minimum, it undergoes a negative decrease, as $\Delta x$ is proportional to the derivative of $E$ with respect to $x$; nevertheless, as the variable approaches the minimum, the changes might make the variable $x$ go from a zone of increase to a zone of decrease, and so to make $x$ oscillate around the limit. In the case of a quadratic function, the first derivative vanishes linearly around the minimum, and therefore these oscillations are damped; in the case of any quadratic function of more variables this effect can be missing: it is thus better to make use of the convergence factor $\eta(k)$. The choice of an opportune $eta(k)$ is the analogue of the variation of the temperature in the Monte-Carlo method, and there exists a theoretical approach for the determination of $\eta(k)$ \cite{ref14}.
\section{More about the Hertz method\label{section5}}
The Hertz methods becomes some more complicated if one wants to obtain a better approximation. In this case, it is better to approximate locally $g(x)$ with linear functions inside the block
\begin{equation}\label{v311}
f(x)=\sum_{\alpha=1}^{\alpha=N}(a^\alpha+b^\alpha x)P_\alpha(x).
\end{equation}
One can imagine that the number of neurons doubles; each neuron becomes a couple in which each element has a different output function.\\
After the scheme Eq. (\ref{v311}), an excellent reconstruction of the function $g(x)=4x(1-x)$ was rendered, as well as of the data pertinent to the ratios of the concentrations of the isotopes of Sn \cite{ref10} considered starting from 540 millions of years ago. It is interesting to note that the plot of the series reconstructed by this method extrapolated at the period before 540 million years is inside the uncertainty of $30\% $ established after the predictions of the ARIMA model.\\
Given any time series which exhibits a stochastic behaviour, it is suited to make use of both approaches, in a way to gain control of the likelihood of the predictions.\\
This method applies also to the distributions of thickness of the rings of the trees \cite{ref15}. Nevertheless, the construction of the blocks $B_\alpha$ was accomplished with good success without working on the data directly, but grouping the data which are obtained from those measured after a transformation, which somehow takes into account the different behaviour of the growth of the trees with respect to their age. It is important to note that the neural approach to this type of problems is made easier according to the fact that there exist on the internet many knots of public domain which make available to the users neural-networks simulators which work in unix environment within the graphic system X-window. These programs can be easily copied by ftp if the computer on which the user works is connected to the web.\\
As an example of neural-network simulators, it is important to consider one constructed by the Hinton group (Toronto), which has many learning methods, i.e. many methods to minimize the cost function $E$ Eq. (\ref{v39}). It is therefore possible to compare many methods of minimization and to chose the most appropriate one as far the current problem is concerned. This simulator contains also a graphic display of the weights which are found, and, therefore, of the solution of the problem of the minimum.The Hertz et al.'s neural network can therefore be used to reconstruct any functional dependence within a time series. The interesting feature of this method is the possibility to accomplish an analysis about the structure and about the distribution of the data. This method seems the best one to address problems of such type, as in comparison with the Kohonen Networks.\\
It is also possible to address this problem by means of a neural network without this introductory analysis about the data distribution. These networks are of common use in many applicative problems , and they are named 'back propagation' networks after the name of the algorithm used within the learning of the network. It is possible to consider the three-layers network: one input layer, an output layer and an intermediate layer. The set of the neural layers and of the connection between the layers forms the architecture of the neural network one is going to investigate next.
\section{More about layers networks\label{section6}} Let $d$ be the characteristic dimension of the time series determined after the methods discussed in the above. It is natural to hypothesize that the input network consist of $d$ neurons, to each of which only one component is presented by the input vectors $\underline{x}\equiv x_1, ..., x_d$. The $d$ input neurons are linked with all the $M$ neurons within the intermediate layer through the weights 
\begin{equation}
W^1_{ij}, \ \ i=1, ..., M, \ \ j=1, ..., d.
\end{equation}
When a $d$-dimensional datum $\underline{x}$ is presented to the input layer, each knot $i$ of the intermediate layer receives a signal spelled after a formula analogous to that of the Hopfield model
\begin{equation}\label{v312}
\sum_{j=1}^{j=d}W^1_{ij}x_j, \ \ i=1, ..., M,
\end{equation}
and, on its turn, the very same neuron emits a signal which is defined after its non-linear output function $z_i$ s.t.
\begin{equation}
z_i=\sigma_1\left(\sum_{j=1}^{j=d}W^1_{ij}x_j\right),
\end{equation}
for which, for instance,
\begin{equation}
    \sigma(x)=\frac{1}{1+e^{-\lambda x}},
\end{equation}
in the cases which are included between $0$ and $1$.\\
The output signals from the $M$ intermediate neurons arrive to the output layer by means of a law analogous to that Eq. (\ref{v312}), such that the generic neuron $i$ from the output layer receives a signal equal to
\begin{equation}\label{v313}
\sum_{j=1}^{j=M}W^2_{ij}z_j.
\end{equation}
The $W^2_{ij}$, $j=1, ..., M$ are the synaptic couplings between the intermediate layer and the output one. The generic output emits a neuron $i$ emits a signal spelled as
\begin{equation}\label{v314}
y_i=\sigma_2\left(\sum_{j=1}^{j=M}W^2_{ij}z_j \right).
\end{equation}
This network has the purpose to admit as an output vector $\underline{y}\equiv y_1, ..., d$ the vector $\underline{x}^{k+1}$ when the date pertinent to the time $k$, i.e. $\underline{x}^k$ is presented to the output layer. It is therefore necessary to determine the set of synaptic links $W^1_{ij}$, $W^2_{lk}$ in a way such that the difference between the output of the network $\underline{y}$ and the wanted output $\underline{x}^{k+1}$ be minimized. It is natural to introduce a cost function of a form
\begin{equation}
E=\sum_{k=1}^{k=N}\sum_{l=1}^{l=d}\left(x_l^{k+1}-\sigma_2(W^2_{lj}\sigma_1(\sum_{m=1}^{m=d}W^1_{jm}x^k_m\right)^2.
\end{equation}
The function $E$ can be minimized again by means of the Monte-Carlo method or by means of the method of the gradient. As it is a complicated function depending on the weights, maxima and minima are expected to be present, such that, in general, there is the need to apply the 'simulated annealing' method \cite{chapIV}. After the analysis of some time series, it was observed that the Hertz et al.'s network and that based on the 'back propagation' provide with the same results as reconstruction; nevertheless, the convergence time of the weights of the latter network is much longer than that of the former network is the case of data issued form systems of stochastic behaviour.
\section{Images-compression methods\label{section7}}
Three images-compression methods will be considered.\\
The Bransley \cite{ref16} method is based on the 'iterated function system' (IFS), which are affine and contractive transformations of the plane.\\
The neural-network-of-'back propagation' method is a method based on a one-intermediate-layer neural network with an input layer and an output layer.\\
The vector-quantization method is realised via Kohonen networks. This approach is very general and applies to the analysis of any time series,i.e.so that it could be treated within the framework of the Hopfield model \cite{sectionIII}. It is possible to reach enormous compression factors via the IFS; neverthelss, the reconstruction mechanism is not clear, and is probably base on the use of other quantization techniques, and on a huge data-base which associates the true image to each compressed image.\\
As an instance to explain the IFS method, it is useful for the sake of simplicity to make use of a 'black and white' image which appears on the screen of a calculator; in other words, a succession of data is considered, $x_1, ..., x_n$, where $x_i$ can acquire the values $0$
 or $1$, and $N$ is the total number of pixels (discrete lattice) which define the screen. The fundamental hypothesis in the Bransley method is that every image can be divided into two different types of subsets, i.e. one in which there is a simple structure whose contours are defined by regular geometric lines, whose approximation by simple function holds. The further method to be considered is one based on the assumption that a complicated structure can be approximated by a fractal set. As instances of fractal structures, the Cantor set and the Serpinski triangle are examples. Almost all of these fractals can be generated by the application of a well-determined family of affine contractive transformations of the plane an infinite number of times.
 \section{More about contractive transformations\label{section8}} A set of contractive transformations $w_1(z), ..., w_k(z)$ of the plane $z=(x, y)\in\mathbb{R}^2$ constitutes a family of IFS maps if the following hypotheses are satisfied:
 \begin{equation}\label{v41}
 w_i(z)\equiv(a_ix+b_iy+c_i, d_ix+e_iy+f_i),\ \ i=1, ..., M,
 \end{equation}
 and
 \begin{equation}
 d(w_i(z_1), w()_i(z_2)\le sd(z_1, z_2),\ \ \forall i=1, ..., k, \\ \forall z_1, z_2\in \mathbb{R}^2, 
 \end{equation}
 with $0<s<1$.\\
 In other words, all the linear transformations which belong to a certain IFS are contractive with the same coefficient of contraction.\\
 Be $W(z)$ the transformation generated by these maps, i.e.
 \begin{equation}\label{v42}
 W(z)=\bigcap_{i=1}^{i=k}w_i(z).
 \end{equation}
 It is straightforward to demonstrate that, if the transformation $W(z)$ is generated by a IFS, by impementing $N$ iterations of the map $W$ starting from a point $z$ chosen at random with unifrm probability in a compact subset $K\in\mathbb{R}^2$, the limit
 \begin{equation}
 lim_{N\rightarrow\infty}(W^o)^N(z)
 \end{equation}
 exists, it does not depend on the choice of $z$ and is a fractal. An example is given after the image which is obtained by applying the following linear transformations \cite{ref16}:
 \begin{subequations}\label{v43}
 \begin{align}
 &x_1=0.5x+1\\
 &y_1=0.5+1\\
 &x_1=0.5x+1\\
 &y_1=0.5y+50\\
 &x_1=0.x+50\\
 &y_1=0.5y+50.
 \end{align}
 \end{subequations}
 $(W^O)^N(z)$ converges with respect to the Hausdorff metric defined in the space of the subsets of a compact $K$ in the plane. Given two subsets $A, B \subset K$, the Hausdorff distance $h_{A, B}$ can be determined in the following way
 \begin{subequations}
 \begin{align}
&h_1(x)=inf_{y\in B}d(x, y),\\
&h_1=sup_{x\in A} h(x),\\
&h_2(y)=inf_{x\in A}d(y, x),\\
&h_2=sup_{y\in B}h_2(y),\\
&h_{A, B}=max(h_1, h_2).
 \end{align}
 \end{subequations}
 As solely images realised on a lattice are here dealt with, the Hausdorff distance can well be substituted by the Hamming distance. In the simple case of black-and-white images, there happens that any two images $A$ and $B$ are two subsets of the rectangular lattice $K$ of dimension $N\times M$ given by the screen, and  it is possible to associate to $A$ ($B$) the vector $\underline{x}_A(i), \ \ i\in K,\ \ (\underline{x}_B, \ \ i\in K)$ such that
 \begin{equation}
 x_A(i)=1 \ \ if\ \ i\in A, x_A(i)=0 \ \ if \ \ i\in K-A.
 \end{equation}
 It is possible to define the distance between the two images $A$ and $B$, $d(A, B)$, thus as the Hamming distance between thee two vectors $\underline{x}_A$ and $\underline{x}_b$ as
 \begin{equation}
 d(A, B)=\frac{1}{N\times M}\sum_{i\in K} \mid \underline{x}_A(i)-\underline{x}_B(i)\mid.
 \end{equation}
 The convergence theorem is valid also for non-discretised figures, as stated in the above. Using the Hamming distance allows one to save a factor $N$ at each step of the procedure (which is demonstrated in the below); indeed, to determine $h_{A, B}$ one needs $N$ operations.\\
 A fractal figure $A$ can be enormously compressed simply after substituting the vector $\underline{x}_A$ which determines it by the coefficients of the transformations $w_1,..., w_M$ which are part of the IFS which generates it. Within this framework, a fractal is defined as any set which can be generated after some IFS. An US trading company founded by Bransley himself elaborated a program, now in commerce, which manages to reduce a picture by Rembrandt in one file of of extension of $50 k$, and, form this, it manages to reconstruct the original picture with a very high precision.\\
 This program apparently solves the following inverse problem: given an image $L=\{ x_1, ..., x_n$, the determination is requested of the transformation generated by an IFS $w_1(z),..., w_M(z)$ as
 \begin{equation}
 W(z)=\cup_{i=1}^{i=k}w_i(z),
 \end{equation}
 which approximates it in the sense that for any point $z$ chosen in the lattice which defines the pixels f the screen of the calculator, one has
 \begin{equation}\label{v44}
 lim_{m\rightarrow\infty}d(W^m(z), L)=0.
 \end{equation}
It is difficult to solve such a problem unless one has a previous catalogue of model-images and of transformations IFS which generate it. There exists a very-well known theorem \cite{ref16} which partially simplifies this the problem: if one manages to determine $\epsilon$ such that $\exists w=\{ w_1, ..., w_k \}$ for which
\begin{equation}
d(w(L), L)\le \epsilon,
\end{equation}
where $w(L)=\cup_{x\in L} w(x)$, then
\begin{equation}
lim_{m\rightarrow\infty} d(w^m(L), L)\le\frac{\epsilon}{1-s};
\end{equation}
 then, the solution of the problem Eq. (\ref{v44}) allows one to approximate at best the given image $L$. Problem Eq. (\ref{v44}) is not trivial; nevertheless, it is possible to try to solve it by Monte-Carlo method. For this, one considers $d(w(L), L)$ as a cost function to be minimised, the states variables are considered as the coefficients $(a_i, b_i, c_i, f_i)_{i=1}^k$ of the IFS $W(w_1, ..., w_k)$; one thereafter discretises the space of coefficients, and, then, starting from a random initial configuration 
 $\underline{\eta}\equiv(a^0_1, b_1^0, ..., f_k^0)$, one introduces the transitions $\underline(\eta)\rightarrow \underline{\eta}_e'=\underline{\eta}_e\pm h$ with probability
 \begin{equation}
 e^{-\beta\Delta}/1+e^{-\beta\Delta}
 \end{equation}
 where
 \begin{equation}
 \Delta=d(w(\underline{\eta'})(L), L)-d(w(\underline{\eta})(L), L)
 \end{equation}
if $\Delta>0$; and with probability 
 \begin{equation}
 \frac{1}{1+e^{-\beta\Delta}}
 \end{equation}
 if $\Delta<0$. This method was applied in the fractal leaf-like structure \cite{ref16}, and to the Serpinski triangle: the results were not satisfying, as it was not possible to reach $\Delta<0.64$ which provides one with a non-suitable reconstruction \cite{ref17}. By dividing the image in square-like boxes of sides $20\times 20$ pixels, and then by using an approach which is more similar to the classical vector-quantization one, and by trying to compress and reconstruct the IFS in each of these blocks, it was possible to obtain a result which was nevertheless not of interest.\\
 The substitution of the Hamming distance with the Hausdorff distance renders the procedure less apt from the point of view of the time requested by the calculator, as the Hausdorff distance requires $2N^2$ operations, but the numerical results on the limiting value of $\Delta$ were of the same type.\\
 Probably the fractal approach is not to be used by itself for each block, but to be associated with other types of compression. As suggested by Hertz \cite{ref13}, it is possible to make use of a network which only one intermediate layer.\\
 It is necessary to consider now a network with $M=2P$ input neurons, $M/2$ intermediate neurons, and $M$ output neurons.\\
 The vector $x_1, ..., x_M$ is given to the input layer, where $M$ is the umber of pixels inside such one block. The datum $x_1$ is presented to the input of the first neuron, the datum $x_2$ is presented to the input of the second neuron, and so on. The neurons from the intermediate layer are connected with all the input neurons and with all the output neurons. Be $z_1, ..., z_[M/2]$ the values of the output of the neurons of the intermediate layer; the following decomposition holds
 \begin{equation}
 z_k=\sigma_1\left( sum_{j=1}^{j=M}w_{kj}^1x_j\right), \ \ k=1, ..., [M/2];
 \end{equation}
 the output is written as
 \begin{equation}
 y_l=\sigma_2\left(w_{lk}^2z_k\right), \ \ l=1, ..., M,
 \end{equation}
 where $w_{lk}^2$ and $w_{kj}^1$ are the synaptic-coupling intensities between the intermediate layer and that of output and between the output state and the intermediate one respectively; and $\sigma_1$ and $\sigma_2$ are the two neural non-linear output functions. It is now the suitable duty to implement the a learning of the net in a way such that, for each black-and-white intensity vector $x_1^\alpha, ..., x_M^\alpha$ , associated to each of the $Q$ blocks $B_\alpha$, $\alpha=1, ..., Q$ in which the image is divided, the output of the net is still $x_1^\alpha, ..., x_M^\alpha$: in this case, it is crucial to outline that the image has been compresses of a factor $2$ because the following step is possible
 \begin{equation}\label{v49}
 \bigcup_{\alpha=1}^{\alpha=Q}(x_1^\alpha, ..., x_M^\alpha)\rightarrow \bigcup_{\alpha=1}^{\alpha=Q}(z_1^\alpha, ..., z_{[M/2]}^\alpha).
 \end{equation}
 It is obtained by minimising the following cost function VEDI
 \begin{equation}\label{v410}
 E=\sum_{\alpha=1}^{\alpha=Q}\sum_{l=1}^{l=M}\left(x_l^\alpha-\sigma_2\left(\sum_{k=1}^{k=[M/2]}w_{lk}^2\sigma_1\sum_{j=1}^{j=M}w_{kj}^1x_j^\alpha\right)\right)^2.
 \end{equation}
 The procedure can be iterated by using a net of the same type, whose input layer is the intermediate layer of the previous one, and an intermediate layer consisting of $[[N/2]2]$ and so on. The reconstruction obtained after the algorithm Eq. (\ref{v49}) allows one to obtain results which are much improved with respect to those obtained within the IFS method; nevertheless, the procedure requests many iterations before an interesting reduction factor is obtained, and it requires again the use of other techniques to get significant results. The problem caused by the repetition of this method is that the number of component of all the vectors $\underline{z}_\alpha$ obtained plus the number of the bytes necessary to individuate the weights of the networks used to complete the 'compressed' vectors $\underline{z}_\alpha$ is equal or greater than the number of bytes that determine the original image. The number id equal in the case of a fractal leaf-like structure.Of course it is possible to try to find a minimum $M$ each time such a network is used.  The most general approach consists in reducing the initial number of bytes which constitutes the image before applying the compression techniques.\\
 The vector quantization is therefore described. The discussion about the problem of the images brought to the introduction of this method, which has nevertheless very general features. An image on a high-resolution screen i seen as a continuous form; this way, the discrete aspect connected with the lattice of pixels which constitute the screen. The idea of quantization consists in the approximation of the color or of the shade of gray in that portion of image in which these features do not have significant variations. According to the previously-accomplished analysis, the image is divided in sets such that, inside each of these sets, there are small variation of intensity of black and white. One associates the mean value of the intensity in the zone with each of these 'small items' of the partition. It is easily shown that this algorithm is a particular case of the following procedure, which is named vector quantization. On starts form observing that, given $a_1,..., x_N$ are the values of the intensities of gray in one of such homogeneous zones, and given $a$ their mean value, $a$
 is the minimum of the following cost function
 \begin{equation}
 E=\sum_{i=1}^{i=N} (x_i-a)^2.
 \end{equation}
 It is now possible to consider a more general situation, in which there is a compact subspace $K\subset \mathcal{R^d}$ and a set of data $\underline{v}\in K$ distributed according to a probability distribution $P(\underline{v})d\underline{v}$. According to this paradigm, it is possible to make a comparison with the case in which there are colors; accordingly, one considers an intensity vector instead of a scalar, each component of which corresponding to the intensity of a determined colour in a certain point of the image. The set $K$ is the set of possible vectors $v$ associated with the image. It is possible to generalise the latter observation by posing the probelm of determining the vectors $\underline{w}_1, ..., \underline{w}_m$ such that the following cost function be continuous
 \begin{equation}
 E=\sum_{i=1}^{i=m}\int_{A}(\underline{w}_i)d\underline{v}_iP(\underline{v})\mid\mid \underline{v}-\underline{w}_i\mid\mid^2
 \end{equation}
 Where the sets $A(\underline{w}_i), i=1, ..., m$ there items of the Vornoi partition of the set $K$, which is defined in the following way:
 \begin{equation}
     A(\underline{w}_i)\equiv\{ \underline{w}\in K : \mid\mid \underline{w}-\underline{w}_i\mid\mid<\mid\mid\underline{w}-\underline{w}_j\mid\mid,\ \ \forall j\neq i\}.
 \end{equation}
It is straightforward to verify that $A(\underline{w}_i)\cap A(\underline{w}_j)=0$ for $i\neq j$, and that $\cup_{i=1}^{i=m}A\underline{w}_i=K$.\\
It is necessary now to apply the method of the gradient to find the minimum of $E$, as
\begin{equation}\label{v412}
\Delta\underline{w}_i=-\eta\int_{A(\underline{w}_i)}d\underline{v}P(\underline{v})(\underline{v}-\underline{v_i}),
\end{equation}
where $\eta$ is an opportune coefficient which varies at the number of iterations of the procedure of updating of the $\underline{w}_i$ varies, where the latter procedure is one of the same type which has been applied in Section \ref{sectionaa}. It is needed to introduce at the present step the formula which approximates the integrals defined on a limited support via the summation of random numbers:
\begin{equation}\label{v413}
\int_Af(x)dx\simeq\sum_if(x_i).
\end{equation}
In Eq. (\ref{v413}), $x_i$ in the summation are random numberrs chosen in the set $A$ with probability $dx$. Byputting togehter Eq. (\ref{v412}) with Eq. (\ref{v413}) and by introducing an index $n$, where the latter is counting the number of iterations, it is possible to write the method of the gradient for $E$ in the following way:
\begin{subequations}\label{v414}
    \begin{align}
    &\underline{w}_i(n+1)=\underline{w}_i(n)-\eta(\underline{v}_n-\underline{w}_i(n))\ \ if\ \ \underline{v}_n\in A(\underline{w}_i(n)),\nonumber\\
    &\underline{w}_j(n+1)=\underline{w}_j(n)\ \ if\ \  \underline{v}(n)\notin A(\underline{w}_j(n))\ \ i, j=1, 2, ..., m.\nonumber
    \end{align}
\end{subequations}
This procedure is easily 'translated' into an algorithm of neural learning. $m$ neurons are considered, and to each neuron $i$ there is initially associated a vector $\underline{w}_i(0)$, where the latter is randomly chosen, i.e. $i=1, ..., m$, and a set of data as that in the above. The $\underline{w}_i$ are named 'weight vectors at the time $n$', and they are equivalent to the synaptic intensities intorduced in the other neural networks. Be $\underline{v}(n)$ a given succession of vectors in $K$, each of which is chosen with probability $P(\underline{v})$, and be $\eta(n)$ a succession of positive constants opportunely chosen. The previous algorithm is equivalent to the following learning rule:
\begin{subequations}\label{v415}
   \begin{align} 
   &\underline{w}_i(n+1)=\underline{w}_i(n)-\eta(n)(\underline{v}_n-\underline{w}_i(n))\ \ \mid\mid \underline{v}(n)-\underline{w}_i(n)\mid\mid<\mid\mid \underline{v}(n)-\underline{w}_j(n)\mid\mid \ \ \forall j\neq i,\nonumber\\
   &\underline{w}_j(n+1)=\underline{w}_j(n) \ \ otherwise, \ \ i, j=1, ..., m.\nonumber
   \end{align}
\end{subequations}
Eq. (\ref{v415}) demonstrates the learning of the Kohonen networks in the particular case where the weight of the neuron $i$ is the only item varying, which satisfies the condition
\begin{equation}\label{eqwin}
    \mid\mid \underline{v}(n)-\underline{w}_i(n)\mid\mid<\mid\mid \underline{v}(n)-\underline{w}_j(n)\mid\mid \ \ \forall j\neq i.
\end{equation}
\section{The winning neuron\label{section9}}
As an insight form Eq. (\ref{eqwin}), the neuron whose weight is qualified by the minimum distance from the vector $\underline{v}$ is named 'the winning neuron'. In the case there should be more than one winning neuron, one neuron is randomly chosen among the winning neurons. This dynamics is a particular case of the Kohonen dynamics, in which the weights of the neurons which belong to a neighbourhood of the winning neuron are varying, too, where the neighbourhood is decreasing with $n$. The set of vectors $\underline{w}_i=lim_{n\rightarrow\infty}\underline{w}_i(n)$ is a generalisation of the average of a homogeneous set of data, and it can be also considered as a set which is obtained by compressing the data which belong to $K$. The compression consists in associating the vector $\underline{w}_i$ to all the vectors which belong to $A(\underline{w}_i)$. The Kohonen network can be used for the recognition of the voice and the limiting weights obtained are the characteristic vectors which represent the phonemes of a language which is being spoken. The recognition is obtained as follows: if a datum $\underline{v}$ is presented to the network, the $\underline{v}$ becomes active only if the neuron which satisfies the condition of the winning neuron: $\underline{v}\in A(\underline{w}_i)$. In \cite{ref14} a theorem is demonstrated, which provides with the sufficient conditions for which the weights $\underline{w}_i(n)$ be convergent in probability to a limiting value. The learning process is a stochastic process, as the samples which are presented to the network are taken with a certain probability distribution, and thus the convergence of the weights to a given limit acquires a meaning only within a probabilistic framework.
\section{Perspectives\label{section10}} There are two conditions requested in \cite{ref14}: the first condition is a certain type of decreasing of $\eta$; the second condition is that the derivative of a certain Ljapunov function, defined on $K^m$, be always negative but except for a limiting value. In other words, there must exist an absolute minimum for the convergence in the probability be accomplished, as it must always happen in the cases when these types of interpretations are applied. Unfortunately, such a Ljapunov function is rather complicated, and there are several local minima which correspond to the metastable states found in \cite{ref18}. Letting the neighbourhood of the winning neuron vary with $n$ allows one to avoid these metastable states; nevertheless, up to now no reason has been justified to demonstrate it. Differently, in \cite{ref14}, an alternative method is proposed, which is based on the analogy of this situation with the research of the absolute minimum for the desordered sysyems.\\
The paper is organised as follows.\\
In Section \ref{sectionaa}, the application of neural networks to reconstruct series of data form non-linear non-deterministic transformations is manifested.\\
In Section \ref{section2}, the Hertz method is surveyed.\\
In Section \ref{section3}, the partition of the phase-space methodology is investigated.\\
In Section \ref{section4}, patterns-learning neural networks are expressed.\\
In Section \ref{section5}, the Hertz method is further studied.\\
In Section \ref{section6}, layers networks are proposed.\\
In Section \ref{section7}, the images-compression methods are investigated.\\
In Section \ref{section8}, the contractive transformations are further scrutinised.\\
In Section \ref{section9}, the qualities of the winner neuron are indicated.\\
In Section \ref{section10}, prospective researches are confronted.

\end{document}